\documentclass[11pt]{article}
\usepackage{latexsym}
\usepackage{hyperref}
\usepackage{amsmath}
\usepackage{amsfonts}
\newcommand{\be}{\begin{equation}}
\newcommand{\ee}{\end{equation}\noindent}
\newcommand{\bear}{\begin{eqnarray}}
\newcommand{\ear}{\end{eqnarray}\noindent}
\newcommand{\no}{\noindent}
\def\unex{\,\e^{(\cdot)}}
\date{}
\renewcommand{\theequation}{\arabic{equation}}

\def\Eins{\mathord{1\hskip -1.5pt
\vrule width .5pt height 7.75pt depth -.2pt \hskip -1.2pt
\vrule width 2.5pt height .3pt depth -.05pt \hskip 1.5pt}}

\newcommand{\slD}{\raise.15ex\hbox{$/$}\kern-.57em\hbox{$D$}}
\newcommand{\slpartial}{\raise.15ex\hbox{$/$}\kern-.57em\hbox{$\partial$}}
\newcommand{\slG}{{{\dot G}\!\!\!\! \raise.15ex\hbox {/}}}

\newcommand{\pol}{\varepsilon}
\def\partder#1#2{{\partial #1\over\partial #2}}

\def\mn{{\mu\nu}}

\def\non{\nonumber}
\def\beqn*{\begin{eqnarray*}}
\def\eqn*{\end{eqnarray*}}

\def\square{\kern1pt\vbox{\hrule height 1.2pt\hbox{\vrule width 1.2pt
   \hskip 3pt\vbox{\vskip 6pt}\hskip 3pt\vrule width 0.6pt}
   \hrule height 0.6pt}\kern1pt}
\def\slash#1{#1\!\!\!\raise.15ex\hbox {/}}

\def\dps{\displaystyle}

\def\half{{1\over 2}}

\def\fourth{{1\over4}}

\def\e{\mbox{e}}

\def\4piTD{{(4\pi T)}^{-{D\over 2}}}
\def\4piT4{{(4\pi T)}^{-2}}

\def\Tintm4{{\dps\int_{0}^{\infty}}{dT\over T}\,e^{-m^2T}
    {(4\pi T)}^{-2}}
\def\Tintm{{\dps\int_{0}^{\infty}}{dT\over T}\,e^{-m^2T}}

\def\tr{{\rm tr}\,}

\def\be{\begin{equation}}\def\ee{\end{equation}}
\def\bea{\begin{eqnarray}}\def\eea{\end{eqnarray}}
\def\ba{\begin{array}}\def\ea{\end{array}}
\def\bea{\begin{eqnarray}}\def\barr{\begin{array}}\def\earr{\end{array}}
\def\eea{\end{eqnarray}}
\begin{document} 
\newcommand{\ho}[1]{$\, ^{#1}$}
\newcommand{\hoch}[1]{$\, ^{#1}$}
\pagestyle{empty}
\renewcommand{\thefootnote}{\fnsymbol{footnote}}

\hfill MaPhy-AvH/2012-08
\vskip .4cm
\begin{center}
{\Large\bf String -- inspired representations of photon/gluon amplitudes}\\
\vskip1.3cm

{\large Naser Ahmadiniaz$^{a,b}$, Christian Schubert$^{a,b,c}$, Victor M. Villanueva$^{a}$
}
\\[1.5ex]

\begin{itemize}
\item [$^a$]
{\it 
Instituto de F\'{\i}sica y Matem\'aticas
\\
Universidad Michoacana de San Nicol\'as de Hidalgo\\
Edificio C-3, Apdo. Postal 2-82\\
C.P. 58040, Morelia, Michoac\'an, M\'exico\\
}
\item [$^b$]
{\it
Dipartimento di Fisica, Universit\`a di Bologna and INFN, Sezione di Bologna,\\
Via Irnerio 46, I-40126 Bologna, Italy
}
\item [$^c$]
{\it 
Institutes of Physics and Mathematics, Humboldt-Universit\"at zu Berlin,\\
Unter den Linden 6, 10099 Berlin, Germany
}
\end{itemize}

\vspace{.1cm}
\vspace{.1cm}

\vskip 2cm
 %
 {\large \bf Abstract}
\end{center}
\begin{quotation}
\noindent
The string-based Bern-Kosower rules provide an efficient way for obtaining
parameter integral representations of the one-loop $N$ - photon/gluon amplitudes
involving a scalar, spinor or gluon loop, starting from a master formula and
using a certain integration-by-parts (``IBP'') procedure. 
Strassler observed that this algorithm also relates to gauge invariance, 
since it leads to the absorption of polarization vectors into field strength tensors.
Here we present a systematic IBP algorithm that works for arbitrary $N$ and leads to an integrand that
is not only suitable for the application of the Bern-Kosower rules but also optimized with respect
to gauge invariance. In the photon case this means manifest transversality at the integrand level,
in the gluon case that a form factor decomposition of the amplitude into transversal and longitudinal parts 
is generated naturally by the IBP, without the necessity to consider  the nonabelian Ward identities.
Our algorithm is valid off-shell, and provides an extremely efficient way of calculating the one-loop
one-particle-irreducible off-shell Green's functions (``vertices'') in QCD. 
It can also be applied essentially unchanged to the one-loop gauge boson amplitudes in open string theory.
In the abelian case, we study the systematics of the IBP also for the practically important case of the one-loop $N$ - photon
amplitudes in a constant field. 

\end{quotation}
\vskip 1cm
\clearpage
\renewcommand{\thefootnote}{\protect\arabic{footnote}}
\pagestyle{plain}

\section{Introduction}
\label{intro}
\renewcommand{\theequation}{1.\arabic{equation}}
\setcounter{equation}{0}

It is by now well-known that techniques originally
developed for the computation of amplitudes in string theory
can be used also for the simplification of calculations in
ordinary quantum field theory.
Already in 1972 Gervais and Neveu observed that the field theory limit of
string theory generates Feynman rules for Yang-Mills theory
in a special gauge that has certain calculational advantages
\cite{gernev}. Actual calculations along these lines were
done, however, only much later 
\cite{grscbr,minahan,kaplunovsky,mettse}.      
A systematic investigation of the field theory limit at the
tree and one-loop level was undertaken by Bern and Kosower, and led to the
establishment of a new set of rules for the construction of the one-loop gluon amplitudes
in QCD \cite{berkos:prl166,berkos:npb362,berkos:npb379}.
These ``Bern-Kosower rules'' were used for a first calculation of
the five -- gluon amplitudes \cite{bediko5glu}.
Their relation to the usual Feynman rules was clarified in \cite{berdun}.

Shortly afterwards, a simpler approach to the derivation of Bern-Kosower type
formulas was initiated by Strassler \cite{strassler1} based on the 
representation of one-loop amplitudes in terms of first-quantized path integrals.
For the case of QED representations of this type
had been known for a long time \cite{feynman:pr80} 
although they had rarely been considered as a tool for state-of-the-art
calculations. 
Subsequently such representations were derived 
also for Yukawa and axial couplings \cite{12,16,dhogag}, and generalized
to higher loop orders \cite{5,15,sascza,daisie} as well as to the inclusion of constant external
fields \cite{shaisultanov,18,40}, gravitation \cite{basvanbook,baszir1,bacozi2,bacozi1,babegi,phograv}
and finite temperature \cite{mckreb:therm,haasch,sato:therm}. 
For a review see \cite{41}.

The central formula in the Bern-Kosower formalism is the
following `master formula':

\begin{eqnarray}
\Gamma_{\rm scal}[k_1,\varepsilon_1;\ldots;k_N,\varepsilon_N]
&=&
{(-ie)}^N
{(2\pi )}^D\delta (\sum k_i)
{\dps\int_{0}^{\infty}}{dT\over T}
{(4\pi T)}^{-{D\over 2}}
e^{-m^2T}
\prod_{i=1}^N \int_0^T 
d\tau_i
\nonumber\\
&&
\!\!\!\!\!\!\!
\times
\exp\biggl\lbrace\sum_{i,j=1}^N 
\Bigl\lbrack  \half G_{Bij} k_i\cdot k_j
-i\dot G_{Bij}\varepsilon_i\cdot k_j
+\half\ddot G_{Bij}\varepsilon_i\cdot\varepsilon_j
\Bigr\rbrack\biggr\rbrace
\mid_{\rm {\rm lin}(\pol_1,\ldots,\pol_N)}
\nonumber\\
\label{scalarqedmaster}
\end{eqnarray}
\no
As it stands, this formula represents the one-loop 
$N$ -- photon amplitude in scalar QED, with photon
momenta $k_i$ and polarisation vectors
$\varepsilon_i$. $m$ denotes the mass, $e$ the charge and $T$ the total proper time
of the scalar loop particle
\footnote{
We work in the Euclidean throughout. With our conventions a Wick rotation 
$k_i^4\rightarrow -ik_i^0, T\rightarrow is$
yields the $N$ - photon amplitude in the
conventions of \cite{weinberg}.}.

Each of the integrals
$\int d\tau_i$ represents one photon leg moving around
the loop. The integrand is written in terms of the
`bosonic' worldline Green's function $G_B$ and its
derivatives,

\begin{eqnarray}
G_B(\tau_1,\tau_2) &=& |\tau_1 - \tau_2|
- {(\tau_1 - \tau_2)^2\over T}\nonumber\\
\dot G_B(\tau_1,\tau_2) &=& {\rm sign}(\tau_1 - \tau_2)
- 2 {{(\tau_1 - \tau_2)}\over T}\nonumber\\
\ddot G_B(\tau_1,\tau_2)
&=& 2 {\delta}(\tau_1 - \tau_2)
- {2\over T}\quad \nonumber\\
\label{GGdGdd}
\end{eqnarray}
\noindent
Dots generally denote a
derivative acting on the first variable,
$\dot G_B(\tau_1,\tau_2) \equiv {\partial\over
{\partial\tau_1}}G_B(\tau_1,\tau_2)$, 
and we abbreviate $G_{Bij}= G_B(\tau_i-\tau_j)$
etc. 
In deriving the master formula a formal exponentiation has been used that needs to be undone
by expanding out the exponential in (\ref{scalarqedmaster}) in the polarization vectors, and keeping
only the terms linear in each of them
(in this paper our interest will be mostly in 
the off-shell case, but it will still be useful to keep the polarization
vectors as book-keeping devices). 

We will not dwell here on the derivation of this formula,
which can be obtained either from the infinite string tension 
limit of string theory
\cite{berkos:prl166,berkos:npb362,berkos:npb379,bern:plb296}
or using the worldline path integral formalism
\cite{strassler1,41}.
Its role in the Bern-Kosower formalism is
central, since it provides the input for the Bern-Kosower rules,
which allow one to obtain, from the scalar QED integrand,
the corresponding integrand for the photon amplitudes
in fermion QED, as well as for the (on-shell) $N$ -- gluon
amplitude in QCD. 
However, those rules do not apply to the master formula
as it stands.
Writing out the exponential in eq.(\ref{scalarqedmaster})
one obtains an integrand

\be \exp\biggl\lbrace 
\cdot
\biggr\rbrace \mid_{\rm multi-linear} \quad={(-i)}^N P_N(\dot G_{Bij},\ddot G_{Bij})
 \exp\biggl[\half \sum_{i,j=1}^N G_{Bij}k_i\cdot
k_j \biggr] \label{defPN} \ee\no 

with a certain polynomial $P_N$
depending on the various  $\dot G_{Bij},\ddot G_{Bij}$ and on the
kinematic invariants.  The application of the Bern - Kosower rules
requires one to now 
remove all second derivatives $\ddot G_{Bij}$ appearing
in $P_N$ by 
suitable integrations by parts in the variables
$\tau_i$.

That this removal of all $\ddot G_B$'s is possible for any $N$ was shown in appendix B
of ~\cite{berkos:npb362}. The new integrand is written
in terms of the $G_{Bij}$'s and $\dot G_{Bij}$'s alone, and
serves as the input for the Bern - Kosower
rules. 
Those allow one to classify the various 
contributions to the $N$ -- photon/gluon amplitude
in terms of $\phi^3$ -- diagrams, and moreover
lead to simple relations between the integrands 
for the scalar, spinor and gluon loop cases.
A complete formulation of the Bern-Kosower
rules is lengthy, and we refer the reader
to \cite{berkos:npb379,bern:plb296,41,berntasi}.
For our present purposes, the most relevant part of the rules is that,
up to a global factor of $-2$ correcting
for the differences in degrees of freedom and
statistics, the integrand for the spinor
loop case can be obtained from the one for
the scalar loop simply
by replacing every closed cycle of $\dot G_B$'s
appearing in $Q_N$ according to the ``replacement rule''

\begin{eqnarray}
\dot G_{Bi_1i_2} 
\dot G_{Bi_2i_3} 
\cdots
\dot G_{Bi_ni_1}
\rightarrow 
\dot G_{Bi_1i_2} 
\dot G_{Bi_2i_3} 
\cdots
\dot G_{Bi_ni_1}
-
G_{Fi_1i_2}
G_{Fi_2i_3}
\cdots
G_{Fi_ni_1}
\nonumber\\
\label{subrule}
\end{eqnarray}
\no
where $G_{F12}\equiv {\rm sign}(\tau_1-\tau_2)$
denotes the `fermionic' worldline Green's
function. Note that an expression is considered a cycle
already if it can be put into cycle form
using the antisymmetry of $\dot G_B$ (e.g.
$\dot G_{B12}\dot G_{B12}=-\dot G_{B12}\dot G_{B21}$).
A similar ``cycle replacement rule''
holds for the gluon loop case. Of course, in the
nonabelian case there will also be many other modifications. 

As was discussed already in \cite{berkos:npb362}, the IBP procedure
is generally ambiguous. However, this does not constitute an impediment
to the application of the Bern-Kosower rules, whose application
requires only that all second derivatives $\ddot G_{Bij}$
have been removed. For this reason,
in the application to the computation
of gluon amplitudes in \cite{berkos:npb379,bediko5glu}
the partial integration had been performed in an essentially arbitrary
way. 

A closer look at the IBP was taken by Strassler in
\cite{strassler2}, who noted that this procedure 
bears an interesting relation to gauge invariance.
For each photon leg, define the corresponding field strength tensor,

\bear
F_i^{\mu\nu}&\equiv&
k_i^{\mu}\varepsilon_i^{\nu}
- \varepsilon_i^{\mu}k_i^{\nu}
\label{defFi}
\ear
Remove all $\ddot G_{Bij}$'s and combine all terms contributing
to a given `$\tau$ - cycle' 
$\dot G_{Bi_1i_2} 
\dot G_{Bi_2i_3} 
\cdots
\dot G_{Bi_ni_1}$.
Then the sum of their Lorentz factors can be written
as a `Lorentz cycle' $Z_n(i_1i_2\ldots i_n)$, defined by

\bear
Z_2(ij)&\equiv&
\half {\rm tr}\Bigl(F_iF_j\Bigr) = \pol_i\cdot k_j\pol_j\cdot k_i - \pol_i\cdot\pol_ik_i\cdot k_j
\non\\
Z_n(i_1i_2\ldots i_n)&\equiv&
{\rm tr}
\Bigl(
\prod_{j=1}^n
F_{i_j}\Bigr) 
\quad (n\geq 3)
\nonumber\\
\label{defZn}
\ear\no
Thus $Z_n$ generalizes the familiar transversal projector.
However, in \cite{strassler2}
no systematic way was found to perform the partial
integrations at arbitrary $N$, and also the absorption of
polarisation vectors into field strength tensors (a process to be called
``covariantization'' in the following) worked only
partially; after the IBP and the sorting of the resulting integrand
in terms of ``cycle content'' some terms are just cycles or products of
cycles, but, starting from the three-point case, there are also
terms with left-overs, called ``tails'' in \cite{strassler2}, and the polarisation
vectors in them were not absorbed yet into field strength tensors. 

The IBP procedure was further studied in \cite{26}, where a definite
partial integration prescription was given which works for any $N$,
preserves the full permutation symmetry and is suitable for computerization.
This algorithm is completely satisfactory as far as 
the application of the Bern-Kosower rules is concerned.
However, it is obviously an interesting question whether 
some IBP algorithm exists which leads
to an integrand where {\sl all} polarisation vectors would
be contained in field strength tensors, thus making gauge invariance, i.e.
transversality, manifest at the integrand level.
It is the purpose of the present work to present such an algorithm
\footnote{An IBP algorithm for the general worldline integrand was
also developed in \cite{bjevan}, but for the unrelated purpose of tensor reduction.}.

In detail, we will do the following:
In chapters \ref{Prep} and \ref{Rrep} we find a surprisingly simple way of using IBPs to
covariantize the Bern-Kosower master formula itself; the resulting representation will
be called the R-representation. In \ref{Qrep}
we summarize the ``symmetric IBP procedure'' of \cite{26}, leading to the
$Q$ - and $Q'$ - representations. The next two
sections \ref{QRrep} and \ref{Srep} define our new algorithm,
which combines elements of both the $R$ -  and the $Q'$ - representation,
and results in what we will call the S-representation of the $N$ photon/gluon amplitudes.
As an aside, in section \ref{2tailalt} we present a further improvement of the 
``two-tail'' which leads to a particularly compact integrand at the four-point level
(but does not seem to generalize to the $N$ - point case). In section \ref{spinor} 
we shortly comment on a direct treatment of the spinor QED case in the
worldline super formalism. Section \ref{nonabelian} is devoted to our main application,
which is the calculation of the one-loop off-shell one-particle-irreducible $N$ - gluon amplitudes (or ``$N$-vertices''). 
While in the abelian case there are never any boundary terms in our IBPs, since all integrations
run over the full loop and the integrand is written in terms of worldline Green's functions with the
appropriate boundary conditions, in the nonabelian case the color ordering of the gluon legs
leads to the restriction of the multiple parameter integrals to ordered sectors, and to the
appearance of such boundary terms \cite{strassler1,strassler2}. Those generally can be combined into
color commutators and, in $x$-space, would in principle allow one to achieve a complete nonabelian extension
of the covariantization, namely to rewrite the final integrand in terms of full nonabelian field strength tensors, and to
complete all derivatives to covariant ones. This is not possible in momentum space, but here instead the IBP procedure
generates a natural form factor decomposition of the $N$-vertices, where the bulk terms are manifestly transversal
and  all non-transversality has been pushed into boundary terms. Finding such a decomposition by standard methods
usually involves a tedious analysis of the nonabelian Ward identities, and so far has been completed only for the three-point
case \cite{balchi2}. 

In section \ref{F} we return to the abelian case, but now with a constant external field added.
Here the replacement rule (\ref{subrule}) applies as well \cite{18,40}, however the IBP procedure is
complicated by the fact that the worldline Green's functions become nontrivial Lorentz matrices.
We will point out here the necessary modifications, and also find a way of using the IBP procedure to
effectively eliminate the nonvanishing coincidence limits of the generalized Green's functions $\dot {\cal G}_B$, ${\cal G}_F$
(see (\ref{exp4}),(\ref{exp6}) below) which otherwise would lead to a proliferation of terms as compared to the
vacuum case. 

In the conclusions section we summarize the properties of our new IBP algorithm,
and discuss various applications, some of which have already been published or are actually in progress.
We shortly comment on possible generalizations to gravity and string theory.

\section{The P - representation}
\label{Prep}
\renewcommand{\theequation}{2.\arabic{equation}}
\setcounter{equation}{0}

We will call ``P-representation" the integrand obtained directly from the expansion of the
Bern-Kosower master formula,

\begin{eqnarray}
\Gamma_{\rm scal}[k_1,\varepsilon_1;\ldots;k_N,\varepsilon_N]
&=&
{(-ie)}^N
{(2\pi )}^D\delta (\sum k_i)
{\dps\int_{0}^{\infty}}{dT\over T}
{(4\pi T)}^{-{D\over 2}}
e^{-m^2T}
\nonumber\\
&&\times
\prod_{i=1}^N \int_0^T 
d\tau_i
P_N(\dot G_{Bij},\ddot G_{Bij})
\exp\biggl\lbrace\sum_{i,j=1}^N 
  \half G_{Bij} k_i\cdot k_j
\biggr\rbrace
\non\\
\label{NointP}
\end{eqnarray}
\no
Explicitly, the polynomial $P_N$ is given by

\bear
P_N
&=&
\dot G_{B1i_1}\varepsilon_1\!\cdot\! k_{i_1}
\dot G_{B2i_2}\varepsilon_2\!\cdot\! k_{i_2}
\cdots 
\dot G_{BNi_N}\varepsilon_N\!\cdot\! k_{i_N}
\non\\ && \hspace{-10pt}
- 
\sum_{{a,b=1}\atop{a<b}}^N
\ddot G_{Bab}\varepsilon_a\!\cdot\!\varepsilon_b\dot G_{B1i_1}\varepsilon_1\!\cdot\! k_{i_1}
\cdots 
\widehat{{\dot G_{Bai_a}\varepsilon_a\!\cdot\! k_{i_a}}}
\cdots
\widehat{\dot G_{Bbi_b}\varepsilon_b\!\cdot\! k_{i_b}}
\cdots 
\dot G_{BNi_N}\varepsilon_N\!\cdot\! k_{i_N}
\non\\&& \hspace{-10pt}
+ 
\sum_{{a,b,c,d=1}\atop{a<b<c<d}}^N
\bigl(\ddot G_{Bab}\varepsilon_a\!\cdot\!\varepsilon_b \ddot G_{Bcd}\varepsilon_c\!\cdot\!\varepsilon_d+\ddot G_{Bac}\varepsilon_a\!\cdot\!\varepsilon_c \ddot G_{Bbd}\varepsilon_b\!\cdot\!\varepsilon_d
+\ddot G_{Bad}\varepsilon_a\!\cdot\!\varepsilon_d \ddot G_{Bbc}\varepsilon_b\!\cdot\!\varepsilon_c\bigr)
\non\\&& \times
\dot G_{B1i_1}\varepsilon_1\!\cdot\! k_{i_1}
\cdots 
\widehat{\dot G_{Bai_a}\varepsilon_a\!\cdot\! k_{i_a}}
\cdots
\widehat{\dot G_{Bbi_b}\varepsilon_b\!\cdot\! k_{i_b}}
\cdots 
\widehat{\dot G_{Bci_c}\varepsilon_c\!\cdot\! k_{i_c}}
\cdots 
\widehat{\dot G_{Bdi_d}\varepsilon_d\!\cdot \!k_{i_d}}
\cdots 
\dot G_{BNi_N}\varepsilon_N\!\cdot \!k_{i_N}
\non\\&&\hspace{-10pt}
- \ldots \non\\
\label{Pnexplicit}
\ear
Here and in the following the dummy indices
$i_1,i_2,\ldots$
should be summed over
from $1$ to $N$, and a `hat' denotes omission.
Note that all terms in $P_N$ are obtained from the first one 
by a simultaneous replacement of pairs of 
$\dot G_{Bri_r}\varepsilon_r\cdot k_{i_r}\dot G_{Bsi_s}\varepsilon_s\cdot k_{i_s}$
by $-\ddot G_{Brs}\varepsilon_r\cdot\varepsilon_s$, which has to be done in all
possible ways.
Note also that $\dot G_{Bii}=0$ by antisymmetry.

These P-representation integrals are still directly 
related to the ones arising in a standard  Feynman or Schwinger parameter calculation
of the $N$ photon amplitude ~\cite{berdun,strassler1}. The exponential
factor will, after a multiple rescaling and performance of the global $T$ --
integration, turn into the standard one-loop  $N$ - point Feynman
denominator polynomial. The $\delta$ - function contained in $\ddot G_{Bij}$ 
will bring together the photons $i$ and $j$, corresponding to a quartic vertex,
and the contributions of such terms match the ones from the
seagull vertex of scalar QED.

\section{The R - representation}
\label{Rrep}
\renewcommand{\theequation}{3.\arabic{equation}}
\setcounter{equation}{0}

Before coming to the ``old'' IBP procedure of \cite{strassler1,26} and its intended improvement,
it will be useful to solve a simpler problem, namely how
to covariantize the Bern-Kosower master formula itself. 
In the following we will often abbreviate

\bear
\e^{\half\sum G_{Bij}k_i\cdot k_j} \equiv \unex
\label{abbunex}
\ear

Consider first the case of $N=2$, where

\bear
P_2 = \dot G_{B12}\pol_1\cdot k_2 \dot G_{B21}\pol_2\cdot k_1 - \ddot G_{B12}\pol_1\cdot \pol_2
\label{P2}
\ear
We choose two vectors $r_1,r_2$ that fulfill 

\bear
r_i\cdot k_i \ne 0
\label{condr}
\ear
but are arbitrary otherwise.
Adding to the integrand $P_2\unex$ the following sum of
total derivative terms,

\bear
&&
-{r_1\cdot \pol_1\over r_1\cdot k_1}
\partial_1\Bigl( \dot G_{B21}\pol_2\cdot k_1\unex\Bigr)
-{r_2\cdot \pol_2\over r_2\cdot k_2}
\partial_2\Bigl( \dot G_{B12}\pol_1\cdot k_2\unex\Bigr)
+
{r_1\cdot \pol_1\over r_1\cdot k_1}
{r_2\cdot \pol_2\over r_2\cdot k_2}
\partial_1\partial_2\unex 
\label{totderP2}
\ear
($\partial_i \equiv \partder{}{\tau_i}$)
the total result is a change of $P_2$ into $R_2$, 

\bear
R_2 := \dot G_{B12}{r_1\cdot F_1\cdot k_2\over r_1\cdot k_1}\dot G_{B21}{r_2\cdot F_2\cdot k_1\over r_2\cdot k_2}
+\ddot G_{B12} 
{r_1\cdot F_1\cdot F_2\cdot r_2\over r_1\cdot k_1r_2\cdot k_2}
\label{R2}
\ear 
Thus we have managed to absorb the polarization vectors into field strength tensors.
And this procedure can be immediately generalized to the $N$-point case:
let us abbreviate

\bear
\rho_i &:=& \frac{r_i\cdot \pol_i}{r_i\cdot k_i} \label{defrho}\\
T_r(i) &:=& \sum_j \dot G_{Bij}{r_i\cdot F_i\cdot k_j\over r_i\cdot k_i} \label{defTr}\\
W_r(ij) &:=& \ddot G_{Bij} 
{r_i\cdot F_i\cdot F_j\cdot r_j\over r_i\cdot k_ik_j\cdot r_j} \label{defW}
\label{defrhoTC}
\ear
and choose vectors $r_1,\ldots,r_N$ fulfilling (\ref{condr}).
Then it is a matter of simple combinatorics to verify that

\bear
P_N\unex + 
\Bigl\lbrack \prod_{a=1}^N (1-\rho_a\partial_a \Delta_a) -1 \Bigr\rbrack
\Bigl\lbrack
P_N\unex 
\Bigr\rbrack
= P_N\Bigl(\dot G_{Bai_a}\varepsilon_a\cdot k_{i_a}\to T_r(a), -\ddot G_{Bab}\pol_a\cdot \pol_b \to W_r(ab)\Bigr) \unex
\non\\
\label{totderPN}
\ear
where the operator $\Delta_a$ is defined as follows: 
each term in $P_N\unex $ either involves the index $a$ in a second derivative factor $\ddot G_{B}$, 
or it carries a factor of $\dot G_{Bai_a}\varepsilon_a\cdot k_{i_a}$. In the former case the term will be annihilated by $\Delta_a$,
in the latter case the action of $\Delta_a$ is to replace the factor of $\dot G_{Bai_a}\varepsilon_a\cdot k_{i_a}$ by $1$.

We can then reexponentiate the new integrand, and arrive at the following covariantized version of the Bern-Kosower
master formula (\ref{scalarqedmaster}):

\begin{eqnarray}
\Gamma_{\rm scal}[k_1,\varepsilon_1;\ldots;k_N,\varepsilon_N]
&=&
{(-ie)}^N
{(2\pi )}^D\delta (\sum k_i)
{\dps\int_{0}^{\infty}}{dT\over T}
{(4\pi T)}^{-{D\over 2}}
e^{-m^2T}
\prod_{i=1}^N \int_0^T 
d\tau_i
\nonumber\\
&&
\hspace{-90pt}
\times
\exp\biggl\lbrace\sum_{i,j=1}^N 
\Bigl\lbrack  \half G_{Bij} k_i\cdot k_j
-i\dot G_{Bij}{r_i\cdot F_i\cdot k_j\over r_i\cdot k_i}
-\half  \ddot G_{Bij} 
{r_i\cdot F_i\cdot F_j\cdot r_j\over r_i\cdot k_i \, r_j\cdot k_j}
\Bigr\rbrack\biggr\rbrace
\Bigg\vert_{{\rm lin}(F_1,\ldots ,F_N)}
\nonumber\\
\label{scalarqedmastercov}
\end{eqnarray}
\no
Thus we have achieved manifest gauge invariance at the integrand level, with a large freedom of
choosing the vectors $r_1,\ldots,r_N$. We will call this the ``R-representation" of the $N$ - photon amplitudes.
Note that it reduces to the original master formula (\ref{scalarqedmaster}) if $r_i\cdot\pol_i=0$ for all $i$.

\section{The $Q$ and $Q'$ - representations}
\label{Qrep}
\renewcommand{\theequation}{4.\arabic{equation}}
\setcounter{equation}{0}

Next, we review the IBP procedure motivated by the Bern-Kosower rules, whose primary purpose is to get rid of all
second derivatives $\ddot G_{B}$ \cite{berkos:npb362,strassler2,26}.

For the two-point case $P_2$ has been written down already in (\ref{P2}).
After an IBP of the second term in either $\tau_1$ or $\tau_2$,
and using $\dot G_{B12} = - \dot G_{B21}$, it turns into

\bear
Q_2 &=& \dot G_{B12}\dot G_{B21}
\bigl(\varepsilon_1\cdot k_2\varepsilon_2\cdot k_1-\varepsilon_1\cdot\pol_2k_1\cdot k_2\bigr)
=
\dot G_{B12}\dot G_{B21}Z_2(12)
\label{Q2}
\ear
Proceeding to the three-point case, here (\ref{Pnexplicit}) becomes

\bear
P_3&=& 
\dot G_{B1i}\varepsilon_1\cdot k_i
\dot G_{B2j}\varepsilon_2\cdot k_j
\dot G_{B3k}\varepsilon_3\cdot k_k
\non\\&&
- 
\Bigl[
\ddot G_{B12}\varepsilon_1\cdot\varepsilon_2
\dot G_{B3i}\varepsilon_3\cdot k_i\, 
+\, (1\rightarrow 2\rightarrow 3)\,
+\, (1\rightarrow 3\rightarrow 2)
\Bigr] 
\non\\
\label{P3}
\ear\no
In this three-point case it is still possible to remove all $\ddot G_B$'s in a single step.
To remove, e. g., the term involving $\ddot G_{B12}\dot G_{B31}$ in the second term of $P_3$, we can add the total derivative term

\bear
-\partial_2 \Bigl(\dot{G}_{B12} \varepsilon_{1}\cdot\varepsilon_{2}\dot{G}_{B31}\varepsilon_{3}\cdot k_{1}
e^{(\cdot)}\Bigr) 
\label{add}
\ear
This term together with five similar ones removes all the $\ddot G_B$'s. Decomposing the
new integrand according to its ``cycle content'',  $P_3$ gets replaced by  $Q_3=Q_3^3+Q_3^2$, where

\bear
Q_{3}^3&=&\dot{G}_{B12}\dot{G}_{B23}\dot{G}_{B31}Z_3(123)  \, ,\non\\
Q_{3}^2&=&\dot{G}_{B12}\dot{G}_{B21}Z_2(12)T(3) +
\dot{G}_{B13}\dot{G}_{B31}Z_2(13)T(2)
+\dot{G}_{B23}\dot{G}_{B32}Z_2(23)T(1) 
\non\\
\label{Q3}
\ear
Note that $Q^3_3$ contains a cycle of length three and $Q_3^2$ a cycle of length two, as indicated by the upper indices,
and that each $\tau$-cycle appears together with the corresponding ``Lorentz-cycle''.
This motivates the further definition of a ``bicycle'' as the product of the two:

\bear
\dot G(i_1i_2\cdots i_n) := \dot G_{Bi_1i_2} 
\dot G_{Bi_2i_3} 
\cdots
\dot G_{Bi_ni_1}
Z_n(i_1i_2\cdots i_n)
\label{defbicycle}
\ear
But the terms of $Q_3^2$ have, apart from the cycle, also a ``one-tail'', defined by

\bear
T(a):=\dot G_{Bai} \varepsilon_a\cdot k_i  
\label{deftail}
\ear 
This tail still has a polarisation vector that is not absorbed into a field strength tensor. It is easy to see that, nonetheless, each term
in $Q_3^2$ is individually gauge invariant; if one replaces in, e.g., the term 

$$  \dot{G}_{B12}\dot{G}_{B21}Z_2(12)\dot{G}_{B3k}\varepsilon_{3}\cdot k_{k}\,\e^{(\cdot )}$$

\noindent
$\varepsilon_3$ by $k_3$, then it becomes proportional to 

$$  \partial_3 \Bigl(  \dot{G}_{B12}\dot{G}_{B21}Z_2(12)\,\e^{(\cdot )}\Bigr)$$

\noindent
However, our aim here is to make gauge invariance manifest even at the integrand level. 
Now in the three-point case there are already various chains of IBP that can be used to remove all the $\ddot G_{B}$'s,
but  if one assumes that the corresponding total derivative terms are added with constant coefficients (i.e., they involve no dependences on 
momentum or polarization other than the ones already present in the term which one wishes to modify), then it is easy to convince oneself that they all
lead to the same $Q_3$ of (\ref{Q3}). 
Thus we have to look for a more general type of IBP. We will now essentially apply the procedure of the
previous section to the tails. 
Consider again the first term in $Q_3^2$ above, eq. (\ref{Q3}).
Choose a momentum vector $r_3$ such that  $r_3\cdot k_3 \ne 0$, and add the total derivative

\bear
- \frac{r_3\cdot\varepsilon_3}{r_3\cdot k_3}Z_2(12)
\partial_3\Bigl(\dot{G}_{B12}\dot{G}_{B21}e^{(\cdot)}\Bigr) 
\label{addtd}
\ear
The addition of this term to the first term in $Q_3^2$, and of similar terms to the second and third one,
transforms $Q_3^2$ into

\bear
R_3^2 &:=& \dot{G}_{B12}\dot{G}_{B21}Z_2(12)\dot{G}_{B3k}\frac{r_3\cdot F_3\cdot k_k}{r_3\cdot k_3}
+
\dot{G}_{B13}\dot{G}_{B31}Z_2(13)\dot{G}_{B2j}\frac{r_2\cdot F_2\cdot k_j}{r_2\cdot k_2}
\nonumber\\
&& + \,\dot{G}_{B23}\dot{G}_{B32}Z_2(23)\dot{G}_{B1i}\frac{r_1\cdot F_1\cdot k_i}{r_1\cdot k_1}
\nonumber\\
\label{defR32}
\ear
Thus we have completed the covariantization of the integrand.  

In the abelian case the 3-point amplitude must, of course, vanish,
which we can see by noting that the integrand is odd
under the orientation-reversing transformation of
variables $\tau_i=T-\tau_i'$,
$i=1,2,3$.

Proceeding to the four-point case, here even using only total derivative terms
with constant coefficients (in the above sense) there are already many ways to remove the
$\ddot G_B$'s by IBP, with a large ambiguity for the final integrand, and it is not obvious how one should proceed.
But certainly one would like to preserve the manifest permutation symmetry of the $N$-photon
amplitudes, and in \cite{26} this was used as a guiding principle to develop the following algorithm:

\begin{enumerate}
\item
In every step, partially integrate away {\sl all} the second derivative factors
$\ddot G_{Bij}$'s appearing in the
term under inspection simultaneously.
This is possible since different $\ddot G_{Bij}$'s never
share variables.

\item
In the first step, for every factor of $\ddot G_{Bij}$ present use
both $\tau_i$ and $\tau_j$ for the IBP, and take the mean of the results.

\item
At every following step, any $\ddot G_{Bij}$
appearing must have been created in the 
previous step. Therefore either both variables $\tau_i$ and $\tau_j$
were used in the previous step, or just
one of them. If both were used, then both should be used again in the
actual IBP step, and the mean of the results be taken.
If only one of the variables was used in the previous step, 
then the other variable should be used in the actual step.

\end{enumerate}

In the four-point case, applying this algorithm to $P_4$ and decomposing the
resulting integrand according to cycle content, 
leads to the following version of the numerator polynomial $Q_4$ \cite{26}:

\bear
Q_4&=&Q_4^4+Q_4^3+Q_4^2-Q_4^{22}\nonumber\\
Q_4^4 &=& 
\dot G(1234)+\dot G(1243) + \dot G(1324)
\non\\
Q_4^3 &=&
\dot G(123)T(4)+\dot G(234)T(1)+\dot G(341)T(2)+\dot G(412)T(3)
\non\\
Q_4^2 &=&
\dot G(12)T(34)+\dot G(13)T(24)+\dot G(14)T(23)+\dot G(23)T(14)+\dot G(24)T(13)+\dot G(34)T(12)
\non\\
Q_4^{22} &=&
\dot G(12)\dot G(34)+\dot G(13)\dot G(24)+\dot G(14)\dot G(23)
\non\\
\label{Q4}
\ear
Here we have now further introduced the two-tail,

\bear
T(ij) &:=&
{\sum_{r,s}}
\biggl\lbrace
\dot G_{Bir}
\varepsilon_i\cdot k_r
\dot G_{Bjs}
\varepsilon_j\cdot k_s
+\half
\dot G_{Bij}
\varepsilon_i\cdot\varepsilon_j
\Bigl[
\dot G_{Bir}k_i\cdot k_r - \dot G_{Bjr}k_j\cdot k_r
\Bigr]
\biggr\rbrace
\label{2tail}
\ear
Thus the final representation of the (still off-shell) four-photon
amplitude in scalar QED becomes

\bear
\Gamma_{\rm scal}
[k_1,\varepsilon_1;\ldots ;k_4,\varepsilon_4]
&=&
{e^4\over (4\pi )^{D\over 2}}
{(2\pi )}^D\delta (\sum k_i)
{\dps\int_{0}^{\infty}}{dT\over T}
T^{-{D\over 2}}
e^{-m^2T}
\non\\
&&\hspace{-30pt}\times
\int_0^T\,d\tau_1\cdots d\tau_4\,
Q_4(\dot G_{Bij})
\exp\biggl\lbrace{1\over 2}\sum_{i,j=1}^4 
G_{Bij} k_i\cdot k_j
\biggr\rbrace\non\\
\label{4photonscal}
\ear\no
We shortly summarize the advantages of this
representation compared to a standard Feynman/Schwinger
parameter integral representation (see \cite{41} for details):

First, the rhs of (\ref{4photonscal}) represents 
already the complete amplitude,
with no need to add ``crossed'' terms. The summation over
``crossed'' diagrams which would have to be done in
a standard field theory calculation here is implicit
in the integration over the various ordered sectors.

Second, the IBP procedure has homogenized the integrand;
every term in $Q_N$ has $N$ factors of $\dot G_{Bij}$
and $N$ factors of external momentum. In the
four-point case this has the additional 
advantage
of making the UV finiteness of the photon-photon
scattering amplitude manifest before integration.
While the original numerator $P_4$ contains terms involving products of two $\ddot G_{Bij}$'s
which lead to spurious divergences in the $T$ - integration, after the IBP
the integrand is finite term by term.

Third, It allows one to obtain the corresponding spinor QED
amplitude by the application of the replacement rule (\ref{subrule}).
In applying the rules it must be observed, though, that the form
of the integrand given aboven still contains, apart from
the explicit cycle factors, additional cycles from
the tail factors for certain values of the dummy indices.
In the four -- point case this occurs
for $Q_4^2$ only: The two -- tails contained in $Q_4^2$ as given in
(\ref{Q4}) above each contain a two-cycle, since the content of the braces on the rhs of (\ref{2tail})
for $r=j, s=i$ turns into 

\bear
\dot G_{Bij}\dot G_{Bji}
\Bigl(
\varepsilon_i\cdot k_j
\varepsilon_j\cdot k_i
-
\varepsilon_i\cdot\varepsilon_j
k_i\cdot k_i
\Bigr)
= 
\dot G(ij)
\label{extract}
\ear
For the application of the ``replacement rules'' it is therefore
convenient to decompose $Q_4$ in a slightly different way
\cite{26}. Namely, note that

\bear
Q_4^2 &=& Q_4^{'2} + 2Q_4^{22}
\label{idQQhat}
\ear
where $Q_4^{'2}$ is obtained from $Q_4^2$ by eliminating the term with
$r=j,s=i$ from the sum over dummy indices. With this
definition, and setting $Q_4^{'(\cdot)} = Q_4^{(\cdot)}$
for the remaining components, we can write

\bear
Q_4 = 
 Q_4^{'4}+ Q_4^{'3}+ Q_4^{'2}+ Q_4^{'22}
\label{rewriteQ4}
\ear
In this form all cycle factors are explicit, and moreover
all the coefficients in the decomposition turn out to be unity.
Generally, we will denote by $T'(i_1\ldots i_n)$ a tail whose cycles have been removed by the
appropriate restrictions on the multiple dummy sums appearing in it.

Thus the four-photon amplitude for the spinor loop case can now be obtained from the scalar
loop formula (\ref{4photonscal}) simply by multiplying with a global factor of $-2$
from spin and statistics, and by replacing, simultaneously, each bicycle $\dot G(i_1\ldots i_n)$
by the corresponding ``super-bicycle''

\bear
\dot G_S(i_1\ldots i_n) := 
(\dot G_{Bi_1i_2} 
\dot G_{Bi_2i_3} 
\cdots
\dot G_{Bi_ni_1}
-
G_{Fi_1i_2}
G_{Fi_2i_3}
\cdots
G_{Fi_ni_1}
)Z_n(i_1\cdots i_n)
\label{defsuperbicycle}
\ear
(the notation refers to the worldline supersymmetry underlying the replacement rule (\ref{subrule}), see \cite{41}).
This ``symmetric partial integration'' procedure has been worked out explicitly for up to the six-photon
case; see \cite{41,26} for the explicit formulas. 

\section{The QR representation}
\label{QRrep}
\renewcommand{\theequation}{5.\arabic{equation}}
\setcounter{equation}{0}

So far we have established two seemingly unrelated IBP procedures, the first
one leading to the manifestly gauge invariant R-representation, the second one to the $Q'$ - representation
that is suitable for the application of the Bern-Kosower rules, but manifestly gauge invariant only in the
cycle factors, not in the tails. We will now combine the two IBP strategies, using the following three simple
observations:

First, it had been noted in the appendix C of \cite{41} that the $Q$ - representation is recursive, in the following sense: Each term
in the cycle decomposition of $Q_N$  contains at least one cycle \cite{41}, so that any tail appearing in the $N$ - point amplitude
has at most $N-2$ arguments. And a tail of length, say, $M$, is related to the (undecomposed) lower-order $Q_M$
simply by writing $Q_M$ in the tail variables, and then extending the range of all dummy variables occurring in it
to run over the full set of indices $1,\ldots, N$. For example, writing out (\ref{2tail}) for the two-tail $T(12)$ in the last term of $Q_4^2$ in (\ref{Q4})
gives

\bear
T(12) = 
\sum_{r,s=1}^4
\dot G_{B1r}
\varepsilon_1\cdot k_r
\dot G_{B2s}
\varepsilon_2\cdot k_s
+
\half
\dot G_{B12}
\varepsilon_1\cdot\varepsilon_2
 \sum_{r=1}^4
\Bigl[
\dot G_{B1r}k_1\cdot k_r - \dot G_{B2r}k_2\cdot k_r
\Bigr]
\label{T12}
\ear
which can also be obtained by writing $Q_2$, defined in (\ref{Q2}), as

\bear
Q_2(12) \equiv 
\dot G_{B12}
\varepsilon_1\cdot k_2
\dot G_{B21}
\varepsilon_2\cdot k_1
+
\half
\dot G_{B12}
\varepsilon_1\cdot\varepsilon_2
\Bigl[
\dot G_{B12}k_1\cdot k_2 - \dot G_{B21}k_2\cdot k_1
\Bigr]
\label{rewriteQ12}
\ear
and introducing appropriate dummy index summations.
That this property holds in general can be easily seen by considering those terms in the cycle factors of the decomposition of $Q_M$
that do not contain factors of $\pol_i\cdot\pol_j$, and thus can have involved IBPs only in the tail and not in the cycle variables;
see the appendix C of \cite{41} for more details.

Second, for each $N$ the symmetric IBP procedure defines a unique $Q_N$ and thus a total derivative term

\bear
S_N\unex \equiv Q_N\unex -P_N \unex
\label{defSN}
\ear
Consider now an arbitrary term in the cycle decomposition of $Q_N$. It will have the form 
$C(i_1\ldots i_L)T(j_1\ldots j_M)$, where $M\leq N-2$, $C(i_1\ldots i_L)$ is a bicycle or product of
bicycles involving the variables $\tau_{i_1},\ldots,\tau_{i_L}$, and $T(j_1\ldots j_M)$ is the unique
(in the symmetric IBP scheme) tail of $M$ variables, written in the remaining variables $\tau_{j_1},\ldots,\tau_{j_M}$.
Consider $C(i_1\ldots i_L)S_M(j_1\ldots j_M)\unex$, where possible dummy variable summations in $S_M$
are extended to run over the full range of variables $\tau_1,\ldots,\tau_N$
as above. This is still a total derivative term (involving only derivatives in
the tail variables), and the above simple relation between the $M$ - tail and $Q_M$ implies, that

\bear
C(i_1\ldots i_L)T(j_1\ldots j_M)\unex - C(i_1\ldots i_L)S_M(j_1\ldots j_M)\unex = C(i_1\ldots i_L)T_p(j_1\ldots j_M) \unex
\label{extendS}
\ear
with a new version $T_p$ of the $M$ - tail which relates to $P_M$ in the same way as 
the standard tail $T$ to $Q_M$,
i.e. by an extension of the dummy index sums.

Continuing with our example above, here we have

\bear
S_2\,\e^{G_{B12}k_1\cdot k_2} &\equiv& (Q_2 - P_2) \,\e^{G_{B12}k_1\cdot k_2} 
\nonumber\\
&=& 
\Bigl\lbrace
\half
\dot G_{B12}
\varepsilon_1\cdot\varepsilon_2
\Bigl[
\dot G_{B12}k_1\cdot k_2 - \dot G_{B21}k_2\cdot k_1
\Bigr]
+ \ddot G_{B12}\pol_1\cdot\pol_2
\Bigr\rbrace
\,\e^{G_{B12}k_1\cdot k_2}
\non\\
&=&
\half (\partial_1-\partial_2) \Bigl(\dot G_{B12}\pol_1\cdot\pol_2 
\,\e^{G_{B12}k_1\cdot k_2}\Bigr)
\non\\
\label{S2}
\ear
and (\ref{extendS}) becomes

\bear
\dot G (34) T(12)\unex - \half (\partial_1-\partial_2) \Bigl(\dot G(34)\dot G_{B12}\pol_1\cdot\pol_2 \unex\Bigr)
=
\dot G(34)T_p(12)\unex
\ear
where $T(12)$ was given in (\ref{T12}) and $T_p(12)$ is given by

\bear
T_p(12)
=
\sum_{r,s=1}^4
\dot G_{B1r}
\varepsilon_1\cdot k_r
\dot G_{B2s}
\varepsilon_2\cdot k_s
-
\ddot G_{B12}
\varepsilon_1\cdot\varepsilon_2
\label{Tp12}
\ear
Finally, the new tail $T_p(\cdots )$ can now be covariantized by a simple extension of (\ref{totderPN}):

\bear
\prod_{a=1}^M (1-\rho_{j_a}\partial_{j_a} \Delta_{j_a})\Bigl(C(\cdot )T_p(j_1\cdots j_M)\unex  \Bigr)
= C(\cdot ) T_p\Bigl(\dot G_{Bai_a}\varepsilon_a\cdot k_{i_a}\to T_r(a), -\ddot G_{Bab}\pol_a\cdot \pol_b \to W_r(ab)\Bigr) \unex
\non\\
\label{totderTPN}
\ear
In our two-tail example, this lead to the following result:

\bear
T_r(12) = 
\dot G_{B1r}{r_1\cdot F_1\cdot k_r\over r_1\cdot k_1}\dot G_{B2s}{r_2\cdot F_2\cdot k_s\over r_2\cdot k_2}
+\ddot G_{B12} 
{r_1\cdot F_1\cdot F_2\cdot r_2\over r_1\cdot k_1k_2\cdot r_2}
\label{2tailr}
\ear
where the sums over $r$ and $s$ run from $1$ to $4$.
This should be compared with $R_2$ of (\ref{R2}).


\section{The S representation}
\label{Srep}
\renewcommand{\theequation}{6.\arabic{equation}}
\setcounter{equation}{0}

Thus in the QR - representation we have the usual decomposition into cycle and tail factors, with the tails
already covariantized, and in a form that generalizes the (lower order) R - representation by an extension of the
dummy index sums to run over all $N$ variables, including those belonging to the cycle factors of the term under consideration.
To finish our quest for an integrand that would be both covariant and suitable for the 
application of the Bern-Kosower rules, two more steps are needed: First, we need to remove
the remaining $\ddot G_B$'s; this can be done by reapplying the symmetric IBP procedure of chapter \ref{Qrep},
without any modifications. And finally all cycles still contained in the tails have to be separated out. 
We will call this final, in some sense  optimized result for the integrand of the $N$ photon/gluon amplitudes, the ``S - representation'',
and denote the corresponding tails by $T'_{\rm s}(\cdot)$. 

Continuing with our example of the two-tail in $Q_4^2$, the first step transforms $T_r(12)$ of (\ref{2tailr}) into

\bear
T_{\rm s}(12) &\equiv&  
\sum_{r,s =1}^4
\dot G_{B1r}{r_1\cdot F_1\cdot k_r\over r_1\cdot k_1}\dot G_{B2s}{r_2\cdot F_2\cdot k_s\over r_2\cdot k_2}
\nonumber\\&&
-\half \dot G_{B12} \Bigl(\sum_{r=1}^4\dot G_{B1r}k_1\cdot k_r - \sum_{s=1}^4\dot G_{B2s}k_2\cdot k_s\Bigr)
{r_1\cdot F_1\cdot F_2\cdot r_2\over r_1\cdot k_1k_2\cdot r_2}
\non\\
\label{2tailqr}
\ear
This form of the two-tail, like the original two-tail $T(12)$ of (\ref{2tail}), still contains a cycle - the terms on the rhs
with $r=2$ and $s=1$ combine to form a $\dot G(12)$, as in (\ref{extract}). Eliminating these terms from the tail one arrives at the final form,

\bear
T'_{\rm s}(12) &\equiv&  
\sum_{{r,s =1}\atop {r,s\ne (2,1)}}^4
\dot G_{B1r}{r_1\cdot F_1\cdot k_r\over r_1\cdot k_1}\dot G_{B2s}{r_2\cdot F_2\cdot k_s\over r_2\cdot k_2}
\nonumber\\&&
-\half \dot G_{B12} \Bigl(\sum_{{r=1}\atop r\ne 2}^4\dot G_{B1r}k_1\cdot k_r - \sum_{{s=1}\atop s\ne 1}^4\dot G_{B2s}k_1\cdot k_s\Bigr)
{r_1\cdot F_1\cdot F_2\cdot r_2\over r_1\cdot k_1k_2\cdot r_2}
\non\\
\label{2tailqrprime}
\ear

For the one-tail there is no difference between $T_r(i)$, $T_{\rm s}(i)$ and $T'_{\rm s}(i)$, being all given by (\ref{defTr}).

We can now write down a covariantized version of the four photon amplitude, simply by taking over (\ref{Q4}), 
(\ref{4photonscal}), and (\ref{rewriteQ4}) and replacing all one-tails by (\ref{defTr}) and all two-tails by (\ref{2tailqrprime}).
Explicit formulas for higher point amplitudes will be given elsewhere.

\section{Alternative version of the two-tail}
\label{2tailalt}
\renewcommand{\theequation}{7.\arabic{equation}}
\setcounter{equation}{0}

For the two-tail, there is actually yet another form which is covariant, free of $\ddot G_B$'s and at the same time
more compact than (\ref{2tailqr}). Starting again with $T_p(12)$ of (\ref{Tp12}) 
we add the following total derivative term to $T_p(12)$ (omitting now the inert cycle factors $C(\cdot )$)

\bear
\frac{1}{(k_1\cdot k_2)^2}\tr (F_1F_2)\,\partial_1\partial_2\unex
+
\frac{1}{k_1\cdot k_2}
\Bigl\lbrack
\varepsilon_1\cdot\varepsilon_2\,\partial_1\partial_2\unex
- \varepsilon_1\cdot k_2\varepsilon_2\cdot k_j\partial_1\Bigl(\dot G_{B2j}\unex\Bigr)
- \varepsilon_2\cdot k_1\varepsilon_1\cdot k_i\partial_2\Bigl(\dot G_{B1i}\unex\Bigr)
\Bigr\rbrack\non\\
\label{totder2alt}
\ear
One obtains the new two-tail

\bear
T_H(12) &\equiv & \dot G_{Bi1}\dot G_{B2j}k_i\cdot H_{12}\cdot k_j
\label{T2modfin}
\ear
where we have introduced the tensor

\bear
H_{12}^{\mu\nu} &\equiv &
\frac{(F_1F_2)^{\mu\nu}k_1\cdot k_2-k_1^{\mu}k_2^{\nu}\tr (F_1F_2)}{(k_1\cdot k_2)^2}
\label{defH}
\ear
Note that $\tr H_{12}=0$ and $H_{12}^T = H_{21}$.
Note also that the term with $i=2,j=1$ in (\ref{T2modfin}) as before produces a $\dot G(12)$, since 

\bear
k_2\cdot H_{12}\cdot k_1 &=&\half \tr (F_1F_2) - \tr (F_1F_2) =- \half \tr (F_1F_2) = -Z_2(12)
\label{HtoZ}
\ear
Thus $T_H(ij)$ can be used as well as $T(ij)$ and $T_{\rm s}(ij)$ in the construction of the four-point amplitudes,
including the application of the replacement rule (\ref{subrule}) and the simple sign change in passing from (\ref{Q4}) to (\ref{rewriteQ4}). 
However, contrary to $T_{\rm s}(ij)$ it appears that $T_H(ij)$ has no natural generalization to the higher-point tails.

\section{The case of spinor QED}
\label{spinor}
\renewcommand{\theequation}{8.\arabic{equation}}
\setcounter{equation}{0}

One of the main purposes of the IBP procedure is to trivialize the transition to the spinor QED case though the
replacement rule (\ref{subrule}). Still, it is interesting to note (and of possible practical relevance for the generalization
to the case of open fermion lines, where the IBP is less attractive due to the existence of boundary terms) that 
in the worldline formalism there is also a more direct treatment of the spin $\half$ case using an approach based on 
explicit worldline supersymmetry \cite{strasslerthesis,15,41,bjevan,babjva}.
It allows one to write down a
master formula for $N$ -- photon scattering
\cite{strasslerthesis}
which is formally analogous
to the one for the scalar loop,
eq.(\ref{scalarqedmaster}):

\begin{eqnarray}
\Gamma_{\rm spin}
[k_1,\varepsilon_1;\ldots;k_N,\varepsilon_N]
&=&
-2
{(-ie)}^N
{(2\pi )}^D\delta (\sum k_i)
{\dps\int_{0}^{\infty}}{dT\over T}
{(4\pi T)}^{-{D\over 2}}e^{-m^2T}
\nonumber\\&&
\!\!\!\!\!\!\!\!\!\!\!\!\!\!\!\!\!\!\!\!
\!\!\!\!\!\!\!\!\!\!\!
\!\!\!\!\!\!\!\!\!\!\!\!\!\!\!\!\!\!\!\!
\!\!\!\!\!\!\!\!\!\!\!\!\!\!\!\!\!\!\!\!
\times
\prod_{i=1}^N \int_0^T 
d\tau_i
\int
d\theta_i
\exp\biggl\lbrace
\sum_{i,j=1}^N
\Biggl\lbrack
\half\hat G_{ij} k_i\cdot k_j
+iD_i\hat G_{ij}\varepsilon_i\cdot k_j
+\half D_iD_j\hat G_{ij}\varepsilon_i\cdot\varepsilon_j\Biggr]
\biggr\rbrace
\Bigg\vert_{{\rm lin}(\pol_1,\ldots ,\pol_N)}
\nonumber\\
\label{supermaster}
\end{eqnarray}
\no
Here we have further introduced integrals
over the Grassmann variables
$\theta_1,\ldots ,\theta_N$, 
such that $\int d\theta_i \theta_i =1$,
and the super derivative

\bear
D &=& {\partial\over{\partial\theta}} - 
   \theta
{\partial\over{\partial\tau}} \label{defD}
\ear
The two worldline Green's functions $G_{B,F}$
now appear combined in the super Green's function

\bear
\hat G(\tau_1,\theta_1;\tau_2,\theta_2)
\equiv G_B(\tau_1,\tau_2) +
\theta_1\theta_2 G_F(\tau_1,\tau_2)
\label{superpropagator}
\ear\no
Now also the polarization vectors 
$\varepsilon_1,\ldots,\varepsilon_N$
are to be treated as Grassmann variables.
The overall sign of the master formula refers to the
standard ordering of the polarization vectors
$\varepsilon_1\varepsilon_2\ldots\varepsilon_N$.

Starting with this master formula, all the manipulations which we have applied in the
previous chapters to the scalar loop integrands can, {\sl mutatis mutandis}, also be
used in the spinor loop case starting from (\ref{supermaster}). Here we will be satisfied with
pointing out that the covariantized Bern-Kosower master formula (\ref{scalarqedmastercov})
generalizes to the spinor QED case as follows:

\begin{eqnarray}
\Gamma_{\rm spin}
[k_1,\varepsilon_1;\ldots;k_N,\varepsilon_N]
&=&
-2 {(-ie)}^N
{(2\pi )}^D\delta (\sum k_i)
{\dps\int_{0}^{\infty}}{dT\over T}
{(4\pi T)}^{-{D\over 2}}
e^{-m^2T}
\non\\&&
\hspace{-140pt}
\times
\prod_{i=1}^N \int_0^T 
d\tau_i
\int
d\theta_i
\exp\biggl\lbrace
\sum_{i,j=1}^N
\Biggl\lbrack
\half\hat G_{ij} k_i\cdot k_j
+iD_i\hat G_{ij}{r_i\cdot F_i\cdot k_j\over r_i\cdot k_i}
-\half D_iD_j\hat G_{ij}
{r_i\cdot F_i\cdot F_j\cdot r_j\over r_i\cdot k_i \, r_j\cdot k_j}
\Biggr]
\biggr\rbrace
\Bigg\vert_{{\rm lin}(F_1,\ldots ,F_N)}
\nonumber\\
\label{spinorqedmastercov}
\end{eqnarray}
\no
where now $F_1,\ldots,F_N$ have to be treated as Grassmann variables.

\section{The nonabelian case}
\label{nonabelian}
\renewcommand{\theequation}{9.\arabic{equation}}
\setcounter{equation}{0}

As far as concerns the calculation of the on-shell QED photon amplitudes, or of the 
on-shell gluon S-matrix elements via the Bern-Kosower rules, the availability of a representation that is manifestly 
transversal at the integrand level is satisfying, but the significance of this fact for
practical calculations is not obvious. To the contrary, it is easy to recognize the 
advantages of such a representation when it comes to the gluon amplitudes off-shell. Here the natural objects 
to consider in QCD are the ``N-vertices'', that is the one-particle-irreducible $N$ - point functions,
and for applications of those it is often essential to decompose them into a basis of transversal and longitudinal tensor structures
(see, e.g., \cite{balchi2,kimbak,corpap,binpap}).
Such a ``transversality-based form factor decomposition'' in the present approach emerges essentially automatically
in the IBP procedure through the appearance of field strength tensors. We have seen how this happens for
the abelian case, but it is true also for the nonabelian case; here in principle one would like to see the full nonabelian
field strength tensor emerging, 

\bear
F_{\mn} \equiv F_{\mn}^a T^a = F^0_{\mn} + ig[A_{\mu}^bT^b,A_{\nu}^cT^c]
\label{defF}
\ear
where by

\bear
F^0_{\mu\nu} \equiv ( \partial_{\mu}A_{\nu}^a - \partial_{\nu}A_{\mu}^a)T^a
\label{deffmn}
\ear
we now denote its ``abelian part''; and indeed Strassler demonstrated already  for some simple cases how this happens \cite{strassler1,strassler2}: 
When the external particles are gluons,
the various ordered sectors of the integral $\int d\tau_1 \cdots d\tau_N$ need to be considered separately, since they carry
different color factors. Therefore boundary terms now arise in the IBP procedure, and the commutator terms are generated as
differences of boundary terms between adjacent sectors that in the abelian case would cancel, but cannot do so any more in the
presence of color since two of the color matrices appear in different orders. In an $x$-space calculation of the effective action,
those commutator terms could then be combined with the ``abelian'' parts of the field strength tensor, but this is not possible
in a momentum space calculation of the $N$ - point function at fixed $N$, since any term in the nonabelian effective action after
Fourier transformation contributes to amplitudes with various numbers of external particles;
e.g., the term $\tr (D_{\mu}F_{\alpha\beta}D^{\mu}F^{\alpha\beta})$ will contribute to the $N$ - point functions with $N$ between two and six.
Generally, each term in the nonabelian effective Lagrangian has a ``core'' term, which has a counterpart already in the abelian
case (in the example this would be $\partial_{\mu}F^0_{\alpha\beta}\partial^{\mu}F^{0\alpha\beta}$) and a number of
``covariantizing'' terms that all involve commutators, and belong to amplitudes with more legs than the core term.
In this section, we will explain the essentials of how to calculate the scalar, spinor and gluon loop contributions to the one-loop $N$ - gluon vertex.
The details and a full recalculation of the three-gluon-vertex will be left to a separate paper \cite{92}.

Starting with the scalar loop case, here the master formula (\ref{scalarqedmaster}) 
 generalizes to the nonabelian case simply by supplying a global color factor, and keeping the gluons in a fixed order:

\begin{eqnarray}
 \Gamma^{a_{1}\dots a_{N}}_{\rm 1PI, scal}[k_{1},\varepsilon_{1};\dots;k_{N},\varepsilon_{N}]
 &=&(-ig)^{N}\mbox{tr}(T^{a_{1}}\dots T^{a_{N}})(2\pi)^{D}i\delta(\sum k_{i})\int_{0}^{\infty} dT(4\pi T)^{-D/2}e^{-m^2 T}\nonumber\\
 && \hspace{-150pt}\times
 \int_{0}^{T}d\tau_{1}\int_0^{\tau_{1}}d\tau_2\dots\int_0^{\tau_{N-2}}d\tau_{N-1}\exp\Bigg\{\sum_{i,j=1}^N\left[\frac{1}{2}
  G_{Bij}k_{i}\cdot k_{j}
-i\dot{G}_{Bij}\varepsilon_{i}\cdot k_{j}+\frac{1}{2}\ddot{G}_{Bij}\varepsilon_{i}\cdot\varepsilon_{j}\right]\Biggr\rbrace
\Bigg\vert_{{\rm lin}(\pol_1,\ldots ,\pol_N)}
\nonumber\\
\label{bknonabelian}
\end{eqnarray}
Here the $T^a$ are the generators of the gauge group in the representation of the loop scalar. 
This treatment of color correponds to a ``color-ordered'' representation (although not necessarily in the usual sense, where
the $T^a$ would be in the fundamental representation, see, e.g., \cite{dixonrev,edinac}). 
Note that we have not only fixed the ordering of the gluons along the loop but also used the translation invariance in $\tau$ 
to set $\tau_{N}=0$. Summation over all $(N-1)!$ inequivalent orderings of the $N$ vertex operators is implied. 
Also, $\Gamma$ has been given an index ``1PI'' to indicate that the rhs gives only the one-particle-irreducible part of the $N$ - gluon
amplitude, not including the reducible part that now also exists, differently from the abelian case. 
Starting from (\ref{bknonabelian}) one can apply the IBP procedure leading from the P - representation to the S - representation as before,
the only novelty being the boundary terms. Since for the bulk terms all the polarization vectors $\pol_i$ get absorbed into the
transversal structures (\ref{defFi}) (which now, however, represent only the ``abelian part'' of the field strength tensor), 
in the final representation the non-transversal
part of the $N$ - vertex must be entirely in the boundary terms, given by lower-point integrals. For those one still has to choose the
vectors $r_i$, preferably in a way that is consistent with the cyclic invariance of the nonabelian amplitudes. In the three-point case
a convenient cyclic choice is $r_1=k_2-k_3$, $r_2=k_3-k_1$, $r_3=k_1-k_2$, and indeed it turns out \cite{92} that, with this choice, the
resulting form factor decomposition matches precisely the standard Ball-Chiu decomposition \cite{balchi2} of the three-gluon-vertex.

Coming to the spinor loop case, here the only issue is whether the replacement rule (\ref{subrule}) can be applied also to all the boundary
terms now arising in the IBP procedure. This is indeed the case, as we can see as follows: it suffices to show the corresponding
statement for the effective action, rather than the momentum space Green's functions. 
Now, the effective Lagrangian can
in principle be written as an infinite series of terms that are Lorentz scalars formed using any number of field strength
tensors and covariant derivatives. As was already mentioned, 
each such term has a core term, whose calculation is not different from the abelian case for either the scalar or spinor loop, 
such that the replacement rule applies to it. All covariantizing terms of a core term must share its coefficient, and low-order calculations show that, as one
would expect, the way this works is that  they all involve the same parameter integral \cite{strassler2,92}. And for the whole structure to continue
to be gauge invariant for the spinor loop case it is necessary that the same replacement rule applies to all the covariantizing terms as well as for the core term.

Similarly one can convince oneself that also the replacement rules that connect the scalar with the gluon loop cases 
\cite{berkos:npb379,bern:plb296,41,strassler1,18} can be extended
from the core terms to the ones involving boundary contributions \cite{92}. 
An additional issue with the gluon loop contribution to the 
$N$ - vertex is that one has to choose a gauge for the gluon propagator. The application of the gluonic replacement
rules gives the $N$ - vertex corresponding to the use of the background field method with Feynman gauge for the quantum part \cite{strassler1,18}, 
which is also known to coincide with the result of the application of the pinch technique \cite{dewedi,hkys,pilaftsis}. 
This version of the gluonic vertex is also the one that leads to SUSY sum rules \cite{binbro}.

\section{The constant external field case}
\renewcommand{\theequation}{10.\arabic{equation}}
\label{F}
\setcounter{equation}{0}

The Bern-Kosower master formula (\ref{scalarqedmaster}) has the following straightforward extension to the $N-$photon amplitude in a constant external field \cite{shaisultanov,18,40}:
\\
\begin{eqnarray}
\Gamma_{\rm scal}[k_{1},\epsilon_{1};\ldots ;k_{N},\epsilon_{N}]&&=(-ie)^{N}(2\pi)^{D}\delta(\sum{k_{i}})
\int^\infty_{0}[4\pi T]^{\frac{-D}{2}}e^{-m^{2}T}\mbox{det}^{-\frac{1}{2}}\left[\frac{\sin \mathcal{Z} }{\mathcal{Z}}\right]\prod_{i=1}^N\int^T_0 d\tau_{i}\nonumber\\
&&\times\exp\bigg\{ \sum_{i,j=1} ^N\left[ \frac{1}{2}k_{i}\cdot\mathcal{G}_{Bij}\cdot k_{j}-i\varepsilon_{i}\cdot\dot{\mathcal{G}}_{Bij}\cdot k_{j}+\frac{1}{2}\varepsilon_{i}\cdot \ddot{G}_{Bij}\cdot
\varepsilon_{j}\right]\bigg\}
\mid_{\rm {\rm lin}(\pol_1,\ldots,\pol_N)}
\nonumber\\
\label{exp1}
\end{eqnarray}
Here we have introduced the matrix $\mathcal{Z}_{\mu\nu}\equiv eTF_{\mu\nu}$ , $\mathcal{G}_{Bij}\equiv\mathcal{G}_{B}(\tau_{i},\tau_{j})$ denotes the ``bosonic"  worldline Green's function
in the constant field background, and $\dot{\mathcal{G}}_{Bij}$,$\ddot{\mathcal{G}}_{Bij}$ its first and second derivatives with respect to $\tau_{i}$.
Explicitly, they are given by \cite{18,40}

 \begin{eqnarray}
&& \mathcal{G}_{B12} =~\frac{T}{2\mathcal{Z}^2}\left(\frac{\mathcal{Z}}{\sin{\mathcal{Z}}}e^{-i\mathcal{Z}\dot{G}_{B12}}+i\mathcal{Z}\dot{G}_{B12}-1\right)\nonumber\\
 &&\dot{\mathcal{G}}_{B12} =~\frac{i}{\mathcal{Z}}\left(\frac{\mathcal{Z}}{\sin{\mathcal{Z}}}e^{-i\mathcal{Z}\dot{G}_{B12}}-1\right)\nonumber\\
 &&\ddot{\mathcal{G}}_{B12}=~2\delta(\tau_{1}-\tau_{2})-\frac{2}{T}\frac{\mathcal{Z}}{\sin{\mathcal{Z}}}e^{-i\mathcal{Z}\dot{G}_{B12}}\nonumber\\
  \nonumber\\
 \label{exp4}
 \end{eqnarray}
while the fermionic Green's function $G_{Fij}$ generalizes to

 \begin{eqnarray}
 \mathcal{G}_{F12} =~G_{F12}\frac{e^{-i\mathcal{Z}\dot{G}_{B12}}}{\cos{\mathcal{Z}}}
 \label{GFcal}
 \end{eqnarray}
These expressions are to be understood as power series in the matrix $\mathcal{Z}$. We note the symmetry properties 
 \newline
 \\
 \bear
 \mathcal{G}_{Bji}=\mathcal{G}^{T}_{Bij},~~~\dot{\mathcal{G}}_{Bji}=-\dot{\mathcal{G}}^{T}_{Bij}, ~~~\ddot{\mathcal{G}}_{Bji}=\ddot{\mathcal{G}}^{T}_{Bij},
 ~~~\mathcal{G}_{Fji}=-\mathcal{G}^{T}_{Fij} 
 \label{exp5}
 \ear
and the coincidence limits

 \begin{eqnarray}
 &&\mathcal{G}_{B}(\tau,\tau)~=~\frac{T}{2\mathcal{Z}^2}\left(\mathcal{Z}\cot \mathcal{Z} -1\right)\nonumber\\
 &&\dot{\mathcal{G}}_{B}(\tau,\tau)~=~i\cot \mathcal{Z}-\frac{i}{\mathcal{Z}}\nonumber\\
 &&\mathcal{G}_{F}(\tau,\tau)~=~-i\tan{\mathcal{Z}}\nonumber\\
 \label{exp6}
 \end{eqnarray}
 Note that those are independent of $\tau$. For the following, it will also be convenient to introduce the ``subtracted" Green's function
 \bear
 \bar{\mathcal{G}}_{B}(\tau_{1},\tau_{2})~:=~\mathcal{G}_{B}(\tau_{1},\tau_{2})-\mathcal{G}_{B}(\tau,\tau)=\frac{T}{2\mathcal{Z}}\left( \frac{e^{-i\mathcal{Z}\dot{G}_{B12}}-\cos{\mathcal{Z}}}{\sin{\mathcal{Z}}}+i\dot{G}_{B12}\right)
 \label{exp7}
 \ear
 The corresponding spinor QED amplitude is obtained from this master formula in the same way as in the vacuum case, with 
the following minor modifications: First, note from (\ref{exp4}),(\ref{exp6}) that 
the worldline Green's functions for the constant field are generally nontrivial Lorentz matrices. Thus the definition of a `cycle' must now be slightly generalized; for example, a term
 $$\varepsilon_{1}\cdot\dot{\mathcal{G}}_{B12}\cdot k_{2}~\varepsilon_{2}\cdot\dot{\mathcal{G}}_{B23}\cdot\varepsilon_{3}~k_{3}\cdot\dot{\mathcal{G}}_{B31}\cdot k_{1}$$
 would have to replaced by 
 $$
 \varepsilon_{1}\cdot\dot{\mathcal{G}}_{B12}\cdot k_{2}~\varepsilon_{2}\cdot\dot{\mathcal{G}}_{B23}\cdot\varepsilon_{3}~k_{3}\cdot\dot{\mathcal{G}}_{B31}\cdot k_{1}~-~\varepsilon_{1}\cdot\mathcal{G}_{F12}\cdot k_{2}~\varepsilon_{2}\cdot\mathcal{G}_{F23}\cdot\varepsilon_{3}~k_{3}\cdot\mathcal{G}_{F31}\cdot k_{1}$$
 
 Second, the above nonvanishing coincidence limits of $\dot{\mathcal{G}}_{B}$ and $\mathcal{G}_{F}$ must be considered
 as cycles of length one, and included in the replacement rule (\ref{subrule}):
 
 \bear
 \dot{\mathcal{G}}_{B}(\tau_{i},\tau_{i})~\to~\dot{\mathcal{G}}_{B}(\tau_{i},\tau_{i})~-~\mathcal{G}_{F}(\tau_{i},\tau_{i})
 \label{exp01}
 \ear
Third, there is now also a replacement rule for the determinant factor:

 \bear
 \mbox{det}^{-\frac{1}{2}}\left[\frac{\sin(\mathcal{Z})}{\mathcal{Z}}\right] ~\to~\mbox{det}^{-\frac{1}{2}}\left[\frac{\tan(\mathcal{Z})}{\mathcal{Z}}\right]
 \label{exp2}
 \ear
This algorithm yields the one-loop $N$~-~photon amplitudes in a constant field for spinor QED.

Although the worldline formalism in a constant field has already found extensive applications 
\cite{mckshe,cadhdu,gussho,17,18,24,korsch,51,52}
a systematic investigation of the IBP procedure for the constant field case has so far been lacking. 
Inspection of the symmetric partial integration algorithm of 
section \ref{Qrep} shows that it applies as well as to this case, and it leads to essentially the same decomposition of the integrand into cycles and tails, with only two modifications: First, one-cycles have to be included; and second, the definition of the ``bicycle of length $n$", now to be denoted by $\dot{\mathcal{G}}({i_{1}i_{2}\cdots i_{n}}),$ has to be changed to
\\
\begin{eqnarray}
\dot{\mathcal{G}}(i)&~:=~&\frac{1}{2}\mbox{tr}(F_{i}\cdot\dot{\mathcal{G}}_{Bii})=\varepsilon_{i}\cdot\dot{\mathcal{G}}_{Bii}\cdot k_{i}\nonumber\\
\dot{\mathcal{G}}(i_{1}i_{2})&~:=~&\frac{1}{2}\mbox{tr}(F_{i_1}\cdot\dot{\mathcal{G}}_{Bi_1 i_2}\cdot F_{i_2}\cdot\dot{\mathcal{G}}_{Bi_2 i_1})\nonumber\\
&~=~&\varepsilon_{1}\cdot\dot{\mathcal{G}}_{B12}\cdot k_2\varepsilon_2 \cdot \dot{\mathcal{G}}_{B21}\cdot k_1-\varepsilon_1\cdot \dot{\mathcal{G}}_{B12}\cdot\varepsilon_2k_2\cdot\dot{\mathcal{G}}_{B21}\cdot k_1\nonumber\\
\dot{\mathcal{G}}(i_1i_2\dots i_{n})&~:=~&\mbox{tr}(F_{i_1}\cdot\dot{\mathcal{G}}_{Bi_1i_2}\cdot F_{i_2}\cdot\dot{\mathcal{G}}_{Bi_2i_3}\cdots F_{i_n}\cdot\dot{\mathcal{G}}_{Bi_ni_1})~~~(n\ge 3)
\label{exp8}
\end{eqnarray}
These bicycles generalize the vacuum ones (\ref{defbicycle}), and will turn into them in the limit $F\to 0$.

However, it turns out that the presence of one-cycles - which would lead to a significant proliferation of terms as compared to the vacuum case - can be altogether avoided, by the following slight modification of the IBP: First, we observe that we have the freedom to shift both the Green's function $\mathcal{G}_{Bij}$ and its derivatives $\dot{\mathcal{G}}_{Bij}$ by arbitrary constant matrices. For $\mathcal{G}_{Bij}$, which appears only in the first term in the exponent of the master formula (\ref{exp1}), this is an obvious consequence of momentum conservation, 
 $\sum_j k_j=0$.
For $\dot{\mathcal{G}}_{Bij}$, we note that it can appear in a $\varepsilon_{i}\cdot\dot{\mathcal{G}}_{Bij}\cdot k_{j}$, $k_{i}\cdot\dot{\mathcal{G}}_{Bij}\cdot k_{j}$ or $\varepsilon_{i}\cdot\dot{\mathcal{G}}_{Bij}\cdot\varepsilon_{j}$. The first type of terms comes directly from the second term in the master formula, and here again a constant matrix added to $\dot{\mathcal{G}}_{Bij}$ drops out immediately
because of momentum conservation. 
The second type of terms arises in the integration-by-parts procedure as a 

$$
\partial_i\sum_{l,j=1}^N\frac{1}{2}G_{Blj}k_l\cdot k_j=\sum_{j=1}^Nk_i\cdot\dot{\mathcal{G}}_{Bij}\cdot k_j
$$

with $j$ running, so that again we can modify $\dot{\mathcal{G}}_{Bij}$ by a constant. The third type arises as the integral of a $\varepsilon_i\cdot\ddot{\mathcal{G}}_{Bij}\cdot\varepsilon_j$ 
in the IBP procedure, and here the constant matrix can be added as an integration constant. 

For the scalar QED case, we can use this freedom to directly eliminate all one-cycles by subtracting from $\dot{\mathcal{G}}_{Bij}$ its (constant) coincidence limit, that is, replacing 
$\dot{\mathcal{G}}_{Bij}$ by
$\bar{\dot{\mathcal{G}}}_{Bij} \equiv \dot{\mathcal{G}}_{Bij}- \dot{\mathcal{G}}_{Bii}$ throughout. 
For the spinor QED case, this would not make sense, since there are still the fermionic one-cycles. Here instead one should anticipate the application of the replacement rule for one-cycles (\ref{exp2}), and use the freedom of modifying 
$\dot{\mathcal{G}}_{Bij}$ to replace it by
\begin{eqnarray}
\hat{\dot{\mathcal{G}}}_{Bij}~&:=&\bar{\dot{\mathcal{G}}}_{Bij}+\mathcal{G}_{Fii}=\dot{\mathcal{G}}_{Bij}-\dot{\mathcal{G}}_{Bii}+\mathcal{G}_{Fii}\nonumber\\
&=&~i\left(\frac{e^{-i\mathcal{Z}\dot{\mathcal{G}}_{B12}}}{\sin{\mathcal{Z}}}-\cot{\mathcal{Z}}-\tan{\mathcal{Z}}\right)
\label{exp9}
\end{eqnarray}
This eliminates all one-cycles, since now by construction
$\hat{\dot{\mathcal{G}}}_{Bii}- \mathcal{G}_{Fii}=0$.

With these modifications, we can write down the one-loop $N$-photon amplitudes in a constant field for scalar and spinor QED
in a way which is almost as compact as in vacuum. For example, the formula (\ref{4photonscal}) for the four-photon amplitude
in scalar QED generalizes to

\begin{eqnarray}
\Gamma_{\rm{scal}}[k_1,\varepsilon_1;\dots;k_4,\varepsilon_4]=\frac{e^4}{(4\pi)^{D/2}}
{(2\pi )}^D\delta (\sum k_i)
\int_0^\infty\frac{dT}{T^{1+D/2}}\e^{-m^2T}\mbox{det}^{-\frac{1}{2}}\left[\frac{\sin{\mathcal{Z}}}{\mathcal{Z}}\right]\nonumber\\
\times\int_0^Td\tau_1d\tau_2 d\tau_3 d\tau_4\left(\mathcal{Q}'^4_4+\mathcal{Q}'^3_4+\mathcal{Q}'^2_4+\mathcal{Q}'^{22}_4\right)\exp{\bigg\{\sum_{i,j=1}^4\left[\frac{1}{2}k_i\cdot \bar{\mathcal{G}}_{Bij}\cdot k_j\right]\bigg\}}\nonumber\\
\label{exp10}
\end{eqnarray}
where
\begin{eqnarray}
\mathcal{Q}'^4_4&=&\bar{\dot{\mathcal{G}}}(1234)+\bar{\dot{\mathcal{G}}}(1243)+\bar{\dot{\mathcal{G}}}(1324)\nonumber\\
\mathcal{Q}'^3_4&=&\bar{\dot{\mathcal{G}}}(123)\bar{\mathcal{T}}'(4)+\bar{\dot{\mathcal{G}}}(234)\bar{\mathcal{T}}'(1)+\bar{\dot{\mathcal{G}}}(341)\bar{\mathcal{T}}'(2)
+\bar{\dot{\mathcal{G}}}(412)\bar{\mathcal{T}}'(3)\nonumber\\
\mathcal{Q}'^2_4&=&\bar{\dot{\mathcal{G}}}(12)\bar{\mathcal{T}}'(34)+\bar{\dot{\mathcal{G}}}(13)\bar{\mathcal{T}}'(24)+\bar{\dot{\mathcal{G}}}(14)\bar{\mathcal{T}}'(23) \nonumber\\
&&+\bar{\dot{\mathcal{G}}}(23)\bar{\mathcal{T}}'(14)+\bar{\dot{\mathcal{G}}}(24)\bar{\mathcal{T}}'(13)+\bar{\dot{\mathcal{G}}}(34)\bar{\mathcal{T}}'(12)\nonumber\\
\mathcal{Q}'^{22}_4&=&\bar{\dot{\mathcal{G}}}(12)\bar{\dot{\mathcal{G}}}(34)+\bar{\dot{\mathcal{G}}}(13)\bar{\dot{\mathcal{G}}}(24)+\bar{\dot{\mathcal{G}}}(14)\bar{\dot{\mathcal{G}}}(23)
\nonumber\\
\label{exp11}
\end{eqnarray}
Here the bicycles $\bar{\dot{\mathcal{G}}}(i_1\cdots i_n)$ differ from the original ones (\ref{exp8}) only by the replacement of $\dot{\mathcal{G}}_{Bij}$ by 
$\bar{\dot{\mathcal{G}}}_{Bij}$, and the tails $\bar{\mathcal{T}}(i_1i_2\dots i_n)$ are isomorphic to the one of the vacuum case (\ref{deftail}),(\ref{2tail}), 
\begin{eqnarray}
\bar{\mathcal{T}}'(i)~&:=&{\sum_{a}}^{'}\varepsilon_i\cdot\bar{\dot{\mathcal{G}}}_{Bia}\cdot k_a\nonumber\\
\bar{\mathcal{T}}'(ij)~&:=&{\sum_{a,b}}^{'}\varepsilon_i\cdot\bar{\dot{\mathcal{G}}}_{Bia}\cdot k_a\varepsilon_j\cdot\bar{\dot{\mathcal{G}}}_{Bjb}\cdot k_b+
\frac{1}{2}{\sum_a}^{'}\varepsilon_i\cdot\bar{\dot{\mathcal{G}}}_{Bij}\cdot\varepsilon_j\left(k_i\cdot\bar{\dot{\mathcal{G}}}_{Bia}\cdot k_a-k_j\cdot\bar{\dot{\mathcal{G}}}_{Bja}\cdot k_a\right)\nonumber\\
\label{exp12}
\end{eqnarray}
As before, a ``prime" on a tail means that its cycle have been removed; note that now this is necessary already for the one-tail.
\\
The corresponding representation for spinor QED is obtained from (\ref{exp10}), (\ref{exp11}) by the change of determinants (\ref{exp2}), multiplication by the global factor of $-2$, and replacement of the bicycles $\bar{\dot{\mathcal{G}}}(i_1\dots i_n)$ by the ``super-bicycles" $\hat{\dot{\mathcal{G}}}_{S}(i_1\dots i_n)$
\begin{eqnarray}
\hat{\dot{\mathcal{G}}}_{S}(i_{1}i_{2})~&:=&\frac{1}{2}\mbox{tr}\left(F_{i_1}\cdot\hat{\dot{\mathcal{G}}}_{B i_{1} i_{2}}\cdot F_{i_2}\cdot\hat{\dot{\mathcal{G}}}_{B i_{2} i_{1}}\right)\nonumber\\
&&-\frac{1}{2}\mbox{tr}\left(F_{i_1}\cdot\mathcal{G}_{Fi_1 i_2}\cdot F_{i_2}\cdot\mathcal{G}_{Fi_2 i_1}\right)\nonumber\\
\hat{\dot{\mathcal{G}}}_{S}(i_{1}i_{2}\cdots i_n)~&:=&\mbox{tr}\left(F_{i_1}\cdot\hat{\dot{\mathcal{G}}}_{B i_{1} i_{2}}\cdot F_{i_2}\cdot\hat{\dot{\mathcal{G}}}_{B i_{2} i_{3}}\cdots F_{i_n}\cdot\hat{\dot{\mathcal{G}}}_{B i_n i_1}\right)\nonumber\\
&&-\mbox{tr}\left(F_{i_1}\cdot\mathcal{G}_{F i_{1} i_{2}}\cdot F_{i_2}\cdot\mathcal{G}_{F i_{2} i_{3}}\cdots F_{i_n}\cdot\mathcal{G}_{F i_n i_1}\right)~~~(n\ge3)\nonumber\\
\label{exp13}
\end{eqnarray}
Moreover, the $\bar{\dot{\mathcal{G}}}_{Bij}$'s in the tails (\ref{exp12}) must also be replaced by $\hat{\dot{\mathcal{G}}}_{Bij}$'s.
\\
The covariantization of the tails does not seem to extend to the constant field case in a natural manner, except for the modification of the 
one-tail (\ref{defrhoTC}),
which generalizes to 
 
 \bear
 T_r(i)~:=\sum_a \frac{r_i\cdot F_i\cdot\bar{\dot{\mathcal{G}}}_{Bia}\cdot k_a}{r_i\cdot k_i}
 \label{exp14}
 \ear
 Also the optimized form of the two-tail (\ref{T2modfin}) does not seem to generalize to the case of the general constant field. 
 It does so, however, for the important special case of a self-dual field, which obeys $F^{2}~\sim~\Eins$ \cite{40,51,52}.
Here one can generalize the total derivative (\ref{totder2alt}) to 

\begin{eqnarray}
\frac{1}{(k_1\cdot k_2)^2}\mbox{tr}(F_1F_2)\partial_1\partial_2 {\rm e}^{(\cdot)}
+\frac{1}{k_1\cdot k_2}\left\lbrack\varepsilon_1\cdot\varepsilon_2\partial_1\partial_2 {\rm e}^{(\cdot)}-\varepsilon_1\cdot k_2\partial_1\left(\varepsilon_2\cdot\bar{\dot{\mathcal{G}}}_{B2j}\cdot k_{j} {\rm e}^{(\cdot)}\right)
-\varepsilon_2\cdot k_1\partial_2\left(\varepsilon_1\cdot\bar{\dot{\mathcal{G}}}_{B1i}\cdot k_{i} \rm{e}^{(\cdot)}\right)\right\rbrack\nonumber\\
\label{exp15}
\end{eqnarray}
and adding this to the (unsubtracted) two-tail $\bar{\mathcal{T}}(12)$ of (\ref{exp12}), one arrives at 
\bear
\mathcal{T}_{H}(ij)~:=~{\sum_{a,b}}k_{a}\cdot\bar{\dot{\mathcal{G}}}_{Bai}\cdot H_{ij}\cdot\bar{\dot{\mathcal{G}}}_{Bjb}\cdot k_b
\label{exp16}
\ear
with the same tensor $H^{\mu\nu}_{ij}$ as in (\ref{defH}).

\section{Conclusions}
\renewcommand{\theequation}{11.\arabic{equation}}
\label{conc}
\setcounter{equation}{0}

We have continued here the
systematic investigation of the Bern-Kosower partial
integration procedure, initiated in \cite{strassler2} and continued in \cite{26}.
We have presented an IBP algorithm that unambiguously leads to a form of the
integrand of the one-loop $N$ photon amplitude in scalar QED which is manifestly
gauge invariant (transversal) at the integrand level, and suitable
for an application of the Bern-Kosower rules. We have worked out this
integrand explicitly at the four-point (two-tail)  level. The $N$ - point integrand contains
$N$ vectors $r_1,\ldots,r_N$ which are constrained only by the condition (\ref{condr}).
Further study will be needed to find out what is the significance of this
ambiguity, and how to make the best use of it. As far as concerns the
on-shell $N$ - photon/gluon amplitudes, one would surmise that the dependence of the
integrand on the vectors $r_i$
is related to the usual dependence on the reference vectors $q_1,\ldots,q_N$ which
one would normally have in the application of the spinor helicity
formalism, but does not exist any more once all polarization vectors are
absorbed into field strength tensors. And indeed, there is clearly a relation:
for example, consider the case of the the on-shell $N$-photon amplitudes with
all helicities positive. When using the $P$ representation together with the
standard spinor helicity formalism (see, e.g., \cite{dixonrev}), one can remove
all terms involving a $\ddot G_{Bij}\pol_i\cdot\pol_j$ by choosing, in the spinor helicity
formalism, the same
reference vector $q_i=q$ for all legs, since then $\pol_i^+\cdot\pol_j^+ = 0$.
Similarly, in the $S$ representation one could make disappear all the factors 
$r_i\cdot F_i\cdot F_j\cdot r_j$ by choosing all $r_i=r$ equal, and $r$ on-shell,
since then 

\bear
r\cdot F_i\cdot F_j\cdot r = \half r \cdot \lbrace F_i,F_j\rbrace \cdot r = -\fourth [ij]^2 r^2 = 0 
\label{killrFFr}
\ear
on account of the identity \cite{56}

\bear
\lbrace F_i^+,F_j^+ \rbrace^{\mu\nu} = -\half [ij]^2\eta^{\mu\nu}
\label{FFid}
\ear
However, it is clear that this match between the freedom of choosing the $r_i$'s and the $q_i$'s
cannot be a perfect one, since the $r_i$'s need not be chosen as on-shell.

All our representations are valid off-shell. This makes them relevant for state-of-the-art
calculations already at the four-point level, since neither the four-photon nor the four-gluon
amplitudes are presently available in the literature fully off-shell (for any spin in the loop). 
This fact is particularly conspicuous in the case of spinor QED, where the on-shell four-photon scattering amplitude 
was obtained already in 1951 by Karplus and Neumann \cite{karneu},
and the extension to the case of two off-shell legs in 1971 by Costantini et al. \cite{cotopi}.
Our integral representations for the QED four-photon amplitudes are, with any of the various definitions
of the tails, manifestly finite term by term and thus suitable for a numerical evaluation as they stand.
For analytical purposes one would still like to reduce the various parameter integrals appearing in them
to scalar box, triangle and bubble integrals. This could be done using existing tensor reduction algorithms
(see \cite{62} and refs. therein), but in a companion paper \cite{4photon} we will rather perform this tensor reduction
in a way that is specifically adapted to the structure of the worldline integrals.

Due to this validity off-shell our representations can also be used to construct, by sewing, all higher-loop
$N$ - photon amplitudes. From the calculation of the two-loop QED $\beta$ function \cite{15} it is 
clear that when calculating those multiloop amplitudes in the worldline formalism significative 
simplifications can be expected from a judicious application of IBP. 

The manifest transversality is probably a more significant issue in the nonabelian case.
Here we have in mind not so much the calculation of on-shell gluon amplitudes, for which other extremely powerful methods have been
developed in recent years (see, e.g., \cite{bediko:annphys,dixonrev2011} and refs. therein), but rather
of the one-particle-irreducible off-shell $N$ - gluon vertices, for which there is presently still a dearth of
efficient methods. For the use of these amplitudes, e.g. in
Schwinger-Dyson equations, it is usually important to have them in a form that separates them into
transversal and non-transversal parts, which normally requires a tedious analysis of the
nonabelian Ward identities. This is one of the reasons why, using standard methods, the
explicit calculation of these off-shell vertices has been completed so far only for the $N=3$ 
case \cite{balchi2,daossa,binbro}. 
In another companion paper \cite{92} 
we use the $S$ representation to recalculate the three-gluon vertex, for the scalar, spinor,
and gluon loops, achieving a drastic reduction in computational effort compared to earlier attempts
(a summary of this calculation was given in \cite{88}).
The main advantages of our approach are that the gluon and spinor loop cases can be effortlessly obtained
from the scalar loop one through the (off-shell extended) Bern-Kosower replacement rules, and that there is no
necessity to solve the Ward identities, rather the decomposition into transversal and non-transversal pieces
emerges automatically in the IBP procedure.
In relation with the latter point it must also be mentioned that the boundary terms 
in the IBP procedure applied to the $N$-vertex always involve color-commutators,
and are always connected to some lower-point term, even appearing with the same integral. 
Thus the algorithm makes it also easy to separate the genuinely new structures appearing in the
$N$-vertex from those that, in terms of the effective action, only serve the completion of lower-point
expressions to fully gauge-invariant ones; see \cite{92}. 

In the abelian case, we have also extended the IBP procedure to the QED $N$-photon amplitudes
in a constant external field. 
Here our main motivation is that these amplitudes can be used for the construction of 
higher-loop Euler-Heisenberg Lagrangians, and those as a tool to obtain insight about the asymptotic behavior of the
QED $N$ - photon amplitudes at high loop orders and large photon number \cite{37,51,52,60,81}.
In that context it would be highly desirable to calculate the three-loop Euler-Heisenberg Lagrangian
in various dimensions,  and indeed the IBP procedure presented in section \ref{F} has made it possible to 
finally achieve this goal at least for the 1+1 dimensional case \cite{85}. 

It should be possible, and very interesting, to generalize our approach to the inclusion of gravity. 
On-shell, the gauge theory Bern-Kosower rules were generalized to the 
construction of string-based representations for the one-loop $N$-graviton amplitudes
in \cite{bedush}, and these gravity rules were then successfully applied at the four-point
level in \cite{dunnor}. They also involve an IBP and ``replacement rules'' connecting the
amplitudes with different spins in the loop.  Worldline path integral representations of the one-loop effective actions
for gravity 
have so far been constructed for spin zero \cite{baszir1}, spin half \cite{bacozi2,bacozi1} and spin one \cite{babegi} in the loop, and  can be used to obtain 
parameter integral representations of the corresponding off-shell one-particle irreducible $N$-graviton amplitudes
that are closely related to those string-based representations. 
They are written in terms of the same worldline Green's functions as the ones for gauge theory,
and the challenge is again to find an IBP algorithm that would allow one to apply the replacement rules and at the 
same time make covariance manifest, where the latter now means the emergence of full Riemann tensors in the IBP.
This algorithm can, however, not be a simple extension of the one which we have presented here, as one can see from
the fact that in gravity there are no boundary terms in the IBP, so that the nonlinear terms in the Riemann tensor now have to be created by $\delta$-functions;
therefore a complete removal of all $\ddot G_B$'s in the IBP is not called for, and in fact also not possible, as one can easily
see already from the case of the graviton propagator.  

Finally, in open string theory the one-loop gauge boson amplitudes can be written in terms of
a master formula that is analogous to the master formula (\ref{scalarqedmaster}),
only that the variables $\tau_i$ parametrize positions along the boundary of the string worldsheet, and the
Green's functions are worldsheet Green's functions (see, e.g., \cite{gswbook}). Since in all of our manipulations we have, apart from 
the translation invariance in proper-time, not used any specific properties of the worldline Green's functions, our
IBP procedure could as well  be applied at the string level to achieve a form factor decomposition based on
gauge invariance. 

\no
{\bf Acknowledgements:} N. A. and C. S. gratefully acknowledge the hospitality of
the Dipartimento di Fisica, Universit\`a di Bologna and INFN, Sezione di Bologna, as well as of
the Institute of Physics and Mathematics of Humboldt University Berlin and of the AEI, Potsdam.
C. S. and V.V. thank CONACYT for financial support through Proyecto CB2008-101353, and N. A. for a PhD fellowship.


\end{document}